\documentclass[12pt,preprint]{aastex} 

\newcommand{\ms}{\mbox{m s$^{-1}~$}}

\newcommand{\msun}{M$_{\odot}$\ }

\newcommand{\mjup}{M$_{\rm JUP}$}

\newcommand{\msini}{$M \sin i~$}

\lefthead{Butler {\it et~al.\/}}
\righthead{HD 83443}
\slugcomment{WARNING: ROUGH DRAFT, Submitted to ApJ}
\received{}
\accepted{}
\begin{document}

\title{On the Double Planet System Around HD 83443$~^{1}$}

\author{R. Paul Butler\altaffilmark{2}, 
Geoffrey W. Marcy\altaffilmark{3},
Steven S. Vogt\altaffilmark{4},
C. G. Tinney\altaffilmark{5},
Hugh R. A. Jones\altaffilmark{6},
Chris McCarthy\altaffilmark{2},
Alan J. Penny\altaffilmark{7},
Kevin Apps\altaffilmark{8},
Brad D. Carter\altaffilmark{9}}

\authoremail{paul@dtm.ciw.edu}

\altaffiltext{1}{Based on observations obtained at 
the Anglo--Australian Telescope, Siding Spring Australia,
and on observations obtained at the W.M. Keck Observatory,
which is operated jointly by the University of California
and the California Institute of Technology.  Keck time has been
granted by both NASA and the University of California.}

\altaffiltext{2}{Department of Terrestrial Magnetism, Carnegie Institution
of Washington, 5241 Broad Branch Road NW, Washington D.C. USA 20015-1305}

\altaffiltext{3}{Department of Astronomy, University of California,
Berkeley, CA USA  94720 and at Department of Physics and Astronomy,
San Francisco State University, San Francisco, CA, USA 94132}

\altaffiltext{4}{UCO/Lick Observatory, 
University of California at Santa Cruz, Santa Cruz, CA, USA 95064}

\altaffiltext{5}{Anglo--Australian Observatory, P.O. Box 296,
Epping, NSW 1710, Australia}

\altaffiltext{6}{Astrophysics Research Institute, Liverpool John Moores
University, Twelve Quays House, Egerton Wharf, Birkenhead CH41 1LD, UK}

\altaffiltext{7}{Rutherford Appleton Laboratory,
Chilton, Didcot, Oxon, OX11 0QX, UK}

\altaffiltext{8}{Physics and Astronomy, University of Sussex,
Falmer, BN1 9QJ, U.K.}

\altaffiltext{9}{Faculty of Sciences, University of Southern Queensland,
Toowoomba, Queensland 4350, Australia}

\begin{abstract}

The Geneva group has reported two Saturn--mass planets orbiting
HD 83443 (K0V) with periods of 2.98 and 29.8 d.  The two planets have
raised interest in their dynamics because of the possible 10:1 orbital
resonance and the strong gravitational interactions.  We report
precise Doppler measurements of HD 83443 obtained with the
Keck/HIRES and the AAT/UCLES spectrometers.  These measurements
strongly confirm the inner planet with period of 2.985 d, with orbital
parameters in very good agreement with those of the Geneva group.
However these Doppler measurements show no evidence of the outer
planet, at thresholds of 1/4 (3 \ms) of the reported velocity
amplitude of 13.8 \ms.  Thus, the existence of the outer planet is in
question.  Indeed, the current Doppler measurements reveal no evidence
of any second planet with periods less than a year.

\end{abstract}


\section{Introduction}
\label{intro}

Several multiple planet systems have been reported, including the triple
planet system around Upsilon Andromedae (Butler et al. 1999) and
double planet systems around GJ876 (Marcy et al. 2001b), HD 83443
(Mayor et al. 2000), HD 168443 (Marcy et al. 2001a, Udry et al. 2002)
and 47 UMa (Fischer et al. 2002).  Double-planet systems have also
been reported in a press release (ESO Press Release 07/01, 2001)
for HD 82943 and HD 74156.  These multiple planet systems contain
planets reported to range from a saturn mass to nearly 10 jupiter
mass, all orbiting within 4 AU.

Interactions between the planets in some of these systems, notably
Gliese 876, are measurable on a time scale of a few years (Lissauer \&
Rivera 2001; Laughlin \& Chambers 2001; Rivera \& Lissauer 2001).
Doppler measurements can reveal the ongoing gravitational
perturbations and constrain both the planet masses and orbital
inclinations.  The interactions and orbital resonances, both
mean--motion and secular, provide clues about the dynamical history of
the systems (Snellgrove et al. 2001; Lee and Peale 2002a; Chiang et
al. 2002).

A most extraordinary double planet system was reported for HD 83443
(Mayor et al. 2000).  Their Doppler measurements made with the CORALIE
spectrometer indicate the existence of two saturn--mass planets that
both reside within 0.2 AU.  The inner planet has an orbital period of
2.985 d, an eccentricity of 0.079 ($\pm$0.033), a minimum (\msini)
mass of 0.34 \mjup, and an orbital distance of 0.038 AU.  The orbital
period is the shortest known for extrasolar planets.  The non--zero
eccentricity of this inner planet is notable as planets with periods
less than 5 d suffer tidal circularization (Wu \& Goldreich 2002).
With the exception of the saturn--mass planet around HD 46375 (Marcy
et al. 2000), all 15 of the previously discovered ``51 Peg--like''
planets have spectral types of G5 or earlier.

Mayor et al. (2000) report a remarkable second planet around HD 83443.
It has an orbital period of 29.83 ($\pm$0.18) d, an eccentricity of
0.42, a minimum (\msini) mass of 0.16 \mjup, and an orbital distance
of 0.17 AU.  This outer planet induces a velocity semiamplitude in
the star of $K$ = 13.8 $\pm$ 1 \ms, rendering it a 14-$\sigma$ detection.
This planet has the smallest \msini yet reported, and is only the third
reported planet with a semiamplitude smaller than 15 \ms (cf. Marcy et al.
2000; Fischer et al. 2002).  

Both of the planets were indicated by Doppler measurements obtained
with the 1.2-m Leonhard Euler telescope at the ESO La Silla
Observatory, which feeds the CORALIE spectrometer (Queloz et
al. 2000a).  Wavelength calibration is achieved by coupling the
telescope and thorium lamp to the spectrometer with a double scrambled
fiber.   The quoted instrumental precision is now 2 \ms (Udry et al. 2002).


As the two planets orbiting HD 83443 are crowded within 0.2 AU, the
system is dynamically active.  Calculations by J.Laskar and W.Benz
(reported in Mayor et al. 2000) and by Wu \& Goldreich (2002) and
Lee \& Peale (2002b) suggest the 
occurrence of significant gravitational interactions between the
two planets.  The tidal circularization time scale for the inner
planet to HD 83443 is estimated to be $3 \times 10^{8}$ yr (Wu \&
Goldreich 2002), while the star is estimated to have an age of 6.5 Gyr.  
In this context, the non--zero eccentricity of the inner planet and the apside
alignment of the two orbits are understood to be due to secular
interactions between the two planets and tidal interactions
with the star (Mayor et al. 2000; Wu \&
Goldreich 2002).  These in turn constrain the orbital inclination of
this system, and the radius of the inner planet (Wu \& Goldreich
2002).  The dynamical evolution that led to the system may involve
migration and resonances (Lee \& Peale 2002b).

Section 2 of this paper will describe new Doppler measurements of 
HD 83443 made from the Keck and AAT telescopes.  Section 3 describes a
search for the two planets, notably the interesting outer planet.  Our
failure to detect the outer planet is discussed in Section 4.

\section{Doppler Velocities and Periodicities}
\label{obs}

HD 83443 (HIP47202) is among the fainter G \& K dwarfs surveyed
by precision Doppler programs with V = 8.23 and B$-$V=0.811
(Perryman et al. 1997), consistent with the assigned
spectral type, K0 V.  The Hipparcos derived distance is 43.5 pc.
(Note that the distance of 23 pc reported in Mayor et al. (2000) is
incorrect.)  The star is photometrically stable at the level of
Hipparcos measurement uncertainty.  The metallicity of the star,
[Fe/H] $=$ $+0.38$ (Santos et al. 2000a), is similar to other stars
with ``51 Peg--like'' planets.  

The precise Doppler observations presented in this Paper
were made with the HIRES echelle spectrometer (Vogt et al. 1994)
on the 10--m Keck I telescope, and the UCLES echelle spectrometer
(Diego et al. 1990) on the 3.9--m Anglo-Australian telescope. 
These spectrometers are operated at a resolution of R $\sim$ 80000
and R $\sim$ 45000 respectively.  Wavelength calibration is
carried out by means of an iodine absorption cell
(Marcy \& Butler 1992) which superimposes a reference iodine
spectrum directly on the stellar spectra (Butler et al. 1996a)
These systems currently achieve photon--limited measurement
precision of 3 \ms.  Detailed information on these two systems,
including demonstration stable stars, can be found in Vogt et al.
2000 (Keck) and Butler et al. 2001 (AAT).

Based on our photometrically estimated metallicity, [Fe/H]=+0.31, we
added HD 83443 to the Anglo--Australian precision Doppler survey in
Feb 1999.  This is among the very faintest stars in the AAT survey.
Ten minute exposures on the 3.9--m Anglo--Australian telescope yield
typical S/N $\sim$70, giving a median measurement uncertainty of 8.0
\ms (Butler et al. 2001).  A total of 16 AAT observations of HD 83443
have been made between Feb 1999 and Mar 2002.  HD 83443 was added to
the Keck precision Doppler survey (Vogt et al. 2000) in Dec 2000 as a
result of the CORALIE announcement of a double planet system.  A total
of 20 Keck observations have been obtained through Mar 2002.  The AAT
and Keck velocity measurements are listed in Table 1.

We fit the velocities with a simple Keplerian model for which the
usual free parameters are $P$, $T_p$, $e$, $\omega$, $K$, as well as a
system velocity zero point, $\gamma$.  Figures 1 and 2 show the Keck
and AAT velocities, respectively, phased at the best--fit Keplerian orbital
period of 2.9856 d.  The reduced $\chi_{\nu}^2$ to the Keplerian fits  
to these data sets are 1.33 and 0.83 respectively.  Figure 3 shows the
combined set of velocities phased.  

Figure 1 shows that a single Keplerian model, without invoking a
second planet, yields a fit to the Keck velocities with an RMS of 4 \ms.
While this strongly confirms the inner planet, the low RMS 
is surprising because
the reported second planet causes a semiamplitude of 13.8 \ms
(Mayor et al. 2000), but is not included in this single--Keplerian
fit.  Similarly the AAT velocities are well fit, within measurement
uncertainty, with a single Keplerian model, as shown in Figure 2.

The combined velocities from Keck and AAT (Figure 3) can also be fit
with a single Keplerian model, and yield a semiamplitude K = 57 \ms,
an orbital eccentricity $e$ = 0.05, and yield a minimum (\msini) mass
of 0.34 \mjup.  Here we have adopted a stellar mass of 0.79 \msun
(Mayor et al. 2000).  The actual stellar mass is probably closer to
1.0 \msun after accounting properly for the high metalicity of the
star.  The RMS to this Keplerian fit for the combined Keck and AAT
observations is 8.1 \ms, and the corresponding reduced $\chi_{\nu}^2$ is
1.39.

Figure 4 shows the residuals to the single Keplerian fit to the
combined Keck and AAT data.  The RMS of the residuals are 3.8 and 10.6 \ms
respectively for the Keck and AAT observations.  The Keplerian orbital
parameters derived from the separate and combined Keck and AAT
observations are listed in Table 2, along with the orbital parameters
from the Geneva web page for both of the planets 83443 announced from
the CORALIE data\footnote{
$http://obswww.unige.ch/{\sim}udry/planet/planet.html$}.  
The orbital parameters for the inner planet, HD 83443b, derived from
the Keck and AAT data sets are in good agreement with the CORALIE result,
differing primarily in that the Keck--AAT orbit is nearly circular, within
measurement uncertainty, as are the other extrasolar planets within
0.05 AU.

The median internal uncertainty of the Keck observations is 2.8 \ms.
Based on the Ca II H \& K lines, we measure the chromospheric
diagnostic $R'HK$ of HD 83443 to be, log(R'$_{\rm HK}$) = -4.85.  The
Doppler velocity ``jitter'' associated with this level of activity for
a K0 V star is 3.0 \ms (Saar et al. 1998; Saar \& Fischer 2000, Santos
et al. 2000b).  Adding the Doppler ``jitter'' in quadrature with the
measurement uncertainty of 3 \ms produces an expected Keplerian RMS to
the Keck data of 4.1 \ms, which is consistent with the observed RMS of
3.8 \ms.

The AAT and Keck data sets have independent and arbitrary velocity
zero--points.  The velocity offset between these two data sets was
thus left as an additional free parameter in the combined Keplerian
fit.  As this velocity zero--point is dependent on the model used to
fit the data, it is not possible to use the combined data set to
search for multiple periodicities.  As the Keck data has both better
phase coverage and significantly higher precision than the AAT data,
we have intensely searched the Keck velocity set for evidence of a
second planet with a period of 29.83 d.  However we also searched
the AAT velocities for the second planet, yielding similar results
as from the Keck data.

Periodogram analysis (Scargle 1982; Gilliland \& Baliunas 1987)
reveals a strong periodicity near 3 days for both the Keck and AAT
data sets.  Figure 5a (top panel) shows the periodogram for the Keck
data.  The highest peak is the 2.986 period.  The dotted line is the
1\% false alarm level.  There remain no other significant peaks,
notably near 29.83 d.  As a strong primary peak can hide secondary
peaks (Butler et al. 1999), we have removed the primary peak by
subtracting off the best--fit Keplerian from Figure 3.  Figure 5b
(lower panel) shows the periodogram of the Keck velocity residuals
from Figure 4.  No significant peaks remain.

We considered the possibility that a 29.8 d periodicity in our
velocities, caused by an outer planet, might have been missed in the
Keck data due to the temporal sampling of velocity measurements.  The
observational window function may cause blind spots at certain
periods.  To test this possibility, we constructed 1000 artificial
velocity sets.  The fake velocities were calculated from the Keplerian
orbital parameters of both planets by simply adding the motion of the
star caused by each planet.  We adopted the orbital parameters for
both planets from Mayor et al. (2000), listed here in Table
2 as planets ``b'' and ``c''.  In the simulation, we sampled the reflex
velocity of the star at the times of the 20 Keck observations listed
in Table 1.  In addition, random noise with an RMS of 4.0 \ms was added
to each of these artificial data set to simulate the combined effects of
Doppler ``jitter'' and the Keck measurement errors.

Each of these fake data sets was then fit with a single least--squares
Keplerian, and the RMS to this single Keplerian fit was recorded.  A
histogram of RMS to the resulting single Keplerian fits is shown in
Figure 6.  The RMS of these fits range from 6.7 to 14.0 \ms.  The
median RMS to the single Keplerian fit is 10.3 \ms.  In contrast,
the RMS of the single Keplerian fit to the actual Keck data
is 3.8 \ms, as indicated by the arrow in Figure 6.  Since none of our
1000 artificial velocity sets could be adequately fit with a single
Keplerian model, the supposed outer planet, if it exists, would similarly
not permit an adequate fit with a single Keplerian model.  Thus there is
less than 0.1\% probability that the outer planet can hide in our actual
velocities.  We conclude that the window function of the Keck observations
would not prevent the detection of the outer planet to HD 83443.  Such an
outer planet, if it existed, should have caused an excess RMS in the velocity 
residuals of $\sim$10 \ms when fit by single Keplerian.  Such velocity
residuals are not seen.

It remains possible that the period of the outer planet of
HD 83443 might be slightly different from that given on the
CORALIE Web page.  If this were so and the window function
of the Keck observations were unfortunately aligned, it might
still be possible that the outer planet could be lurking in
the Keck data set.  To test this, we have fit the Keck data
set with a double Keplerian, using as the input guess the
double Keplerian parameters from the CORALIE web page (listed
in Table 2).  The period, eccentricity, and velocity semiamplitude of
the inner and outer inner planets have been frozen at the reported values
of the supposed outer planet, but the
remaining Keplerian parameters including time of periastron and
$\omega$ have been allowed to float.  Outer planet periods ranging
from 28 to 32 d were systematically attempted in steps of 0.001 d.
Figure 7 shows the resulting best--fit reduced $\chi_{\nu}^2$ for
each of the trial periods.  The arrow indicates the location of
the 29.83 d period, which yields best--fit reduced $\chi_{\nu}^2$
of 2.31, much worse than the single Keplerian fit to the Keck
data set with reduced $\chi_{\nu}^2$ of 1.33.

To estimate the largest semiamplitude allowed by the Keck velocities for
a planet in a $\sim$30 d orbit, we again fit the Keck data with a
double Keplerian as in Figure 7, but this time allowed the velocity
semiamplitude of the outer planet to float. 
Figure 8 shows this best--fit semiamplitude for
a potential outer planet having orbital periods ranging from 28 to 32 d.
At 29.83 d, the best--fit amplitude is 2.8 \ms.  The Keck data rule out
any periodicities between 29 and 31 days with an amplitude greater
than 3 \ms.  Given the temporal sampling of the Keck data, it remains
possible to hide a 13.8 \ms semiamplitude with a period near 27.7 d
from the current Keck data set.  This is $\sim$10 $\sigma$ removed from
the CORALIE period of 29.83 d.

\section{Discussion}

Precision Doppler observations made with the 10--m Keck and 
the 3.9--m AAT strongly confirm the existence of the inner
planet orbiting HD 83443, and indicate the orbital parameters
are in very good agreement with those reported by Mayor et al. (2000).   
The present orbital parameters differ only marginally
in that the orbit of the inner planet is circular within
measurement uncertainty for the Keck and AAT data,
similar to the other known close--in planets.

But our Doppler measurements did not detect the 29.8 d outer planet,
despite the clear ability to do so.
The present measurements impose a limit on any such 
velocity periodicity at a level of no more than 3 \ms, well below the reported
velocity amplitude of 13.8 \ms.  Orbital periods within 2 days
of 29.8 d would have been detected.  The supposed velocity amplitude
of 13.8 \ms is four times larger than the uncertainties in our
velocity measurements rendering the outer planet immediately
detectable.  Various tests suggest quantitatively that the velocities
should have revealed the outer planet.  The velocities from the Keck
and AAT telescopes could have independently detected the outer planet,
but neither data set revealed it.

We considered various possible reasons that we failed to detect 
the outer planet.  One possibility is that some interactive
resonance between the two planets causes the reflex
velocity of the star to mimic insidiously a single Keplerian orbit.
That is, perhaps the 10:1 ratio of orbital periods, along with
gravitational interactions, yields a final reflex velocity that traces
a single Keplerian velocity curve.  If so, we might be fooled into
fitting the velocities with such a simple model.  We find this possibility
unlikely.  As shown by W. Benz (Mayor et al. 2000) and by Wu \& Goldreich
(2002), the gravitational interactions yield temporal evolution of the orbits 
on a time scale of $\sim$1000 yr rather than a few years.  Thus we expect
the outer planet, if it exists, to remain in a coherent orbit during the
few year duration of the present observations.  Moreover, the 10:1 ratio
of the two periods do not constitute a powerful Fourier harmonic from which
a single Keplerian may be constructed (as is the case with a 2:1 ratio
of periods).

We remain puzzled by the discrepancy between the reported CORALIE
results and the velocities we have obtained with Keck/HIRES and
AAT/UCLES.

Of the 75 extrasolar planet candidates (\msini $<$ 13 \mjup)
announced from precision Doppler
surveys\footnote{$http://exoplanets.org/almanacframe.html$},
a total of 57 have been published in refereed journals.
These planets are listed in Table 3, which also notes the telescope
from which the data originates, and whether the actual Doppler
velocities are publicly available.
Refereed precision velocity confirmations are also included.
Substellar candidates found by other techniques such as astrometry
and low precision Doppler velocities are not included.
An additional 4 planet candidates have been announced in conference
proceedings\footnote{HD 6434, HD 19994, HD 121504, HD 190228}
(Queloz et al. 2000b; Sivan et al. 2000).  Doppler velocity
measurements are not available for candidates that have only
been published in conference proceedings.  An additional 12
claimed Doppler planets, all of which were announced more than
1 year ago, have not been submitted to either a conference
proceeding or a refereed journal.

While the discovery of extrasolar planets has become seemingly
commonplace over the past 6 years, we still consider the detection
of planets orbiting other stars as extraordinary, and as such
worthy of the dictum, ``Extraordinary claims require
extraordinary evidence.''  Publishing discovery data in a
refereed journal remains a crucial part of the process,
though this is not in itself sufficient to establish the
credibility of a planet claim.  It remains extremely likely
that at least a handful of the reported planets 
do not in fact exist.
Multiple confirmation both by independent precision Doppler
teams and by completely independent techniques remain the 
only means by which to ensure the veracity of extrasolar
planet claims.

\acknowledgements

We acknowledge support by NSF grant AST-9988087, NASA grant
NAG5-12182, and travel support from the Carnegie Institution
of Washington (to RPB), NASA grant NAG5-8299 and NSF grant
AST95-20443 (to GWM), NSF grant AST-9619418 and NASA grant
NAG5-4445 (to SSV), and by Sun Microsystems.  We thank the NASA
and UC Telescope assignment committees for allocations of Keck
telescope time, and the Australian (ATAC) and UK (PATT) Telescope
assignment committees for allocations of AAT time.  We thank
Debra Fischer, Greg Laughlin, Doug Lin, Stan Peale and Man Hoi Lee
for valuable conversations.  The authors wish to extend special thanks
to those of Hawaiian ancestry on whose sacred mountain of Mauna Kea
we are privileged to be guests.  Without their generous hospitality,
the Keck observations presented herein would not have been possible.

\clearpage

\begin{figure}
\centerline{\scalebox{.75}{\rotatebox{90}{\includegraphics{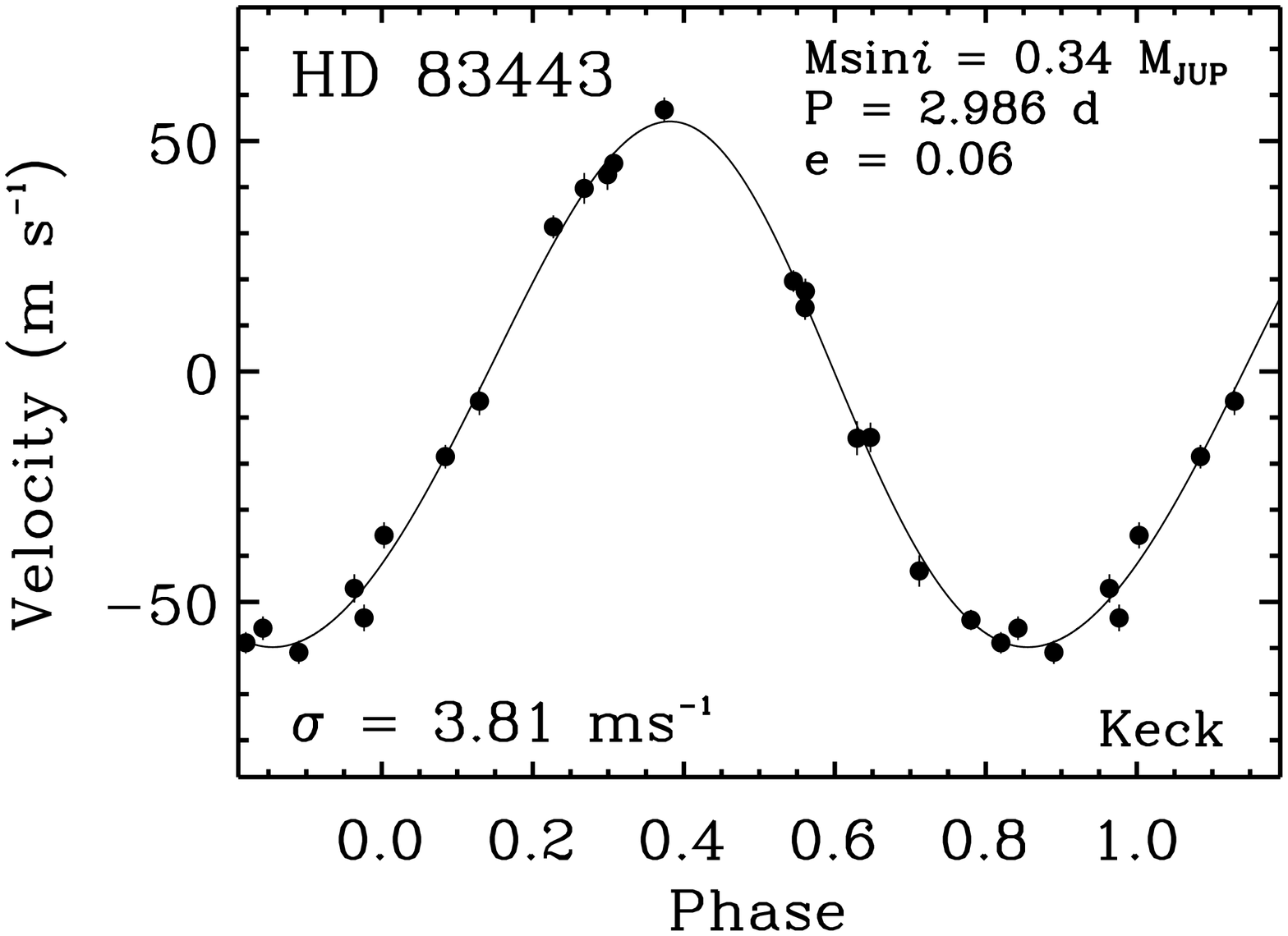}}}}
\caption{Phased Doppler velocities for HD 83443 from Keck.  The solid
line is the best--fit Keplerian orbit assuming only a single planet.
The period, p = 2.986 d, and semiamplitude, K = 57 \ms, are nearly
identical to the CORALIE parameters for the inner planet.  The small RMS
of the residuals of 3.8 \ms is consistent with errors, implying
no evidence for a second planet.}
\label{Keck_Phased_Velocities}
\end{figure}

\begin{figure}
\centerline{\scalebox{.75}{\rotatebox{90}{\includegraphics{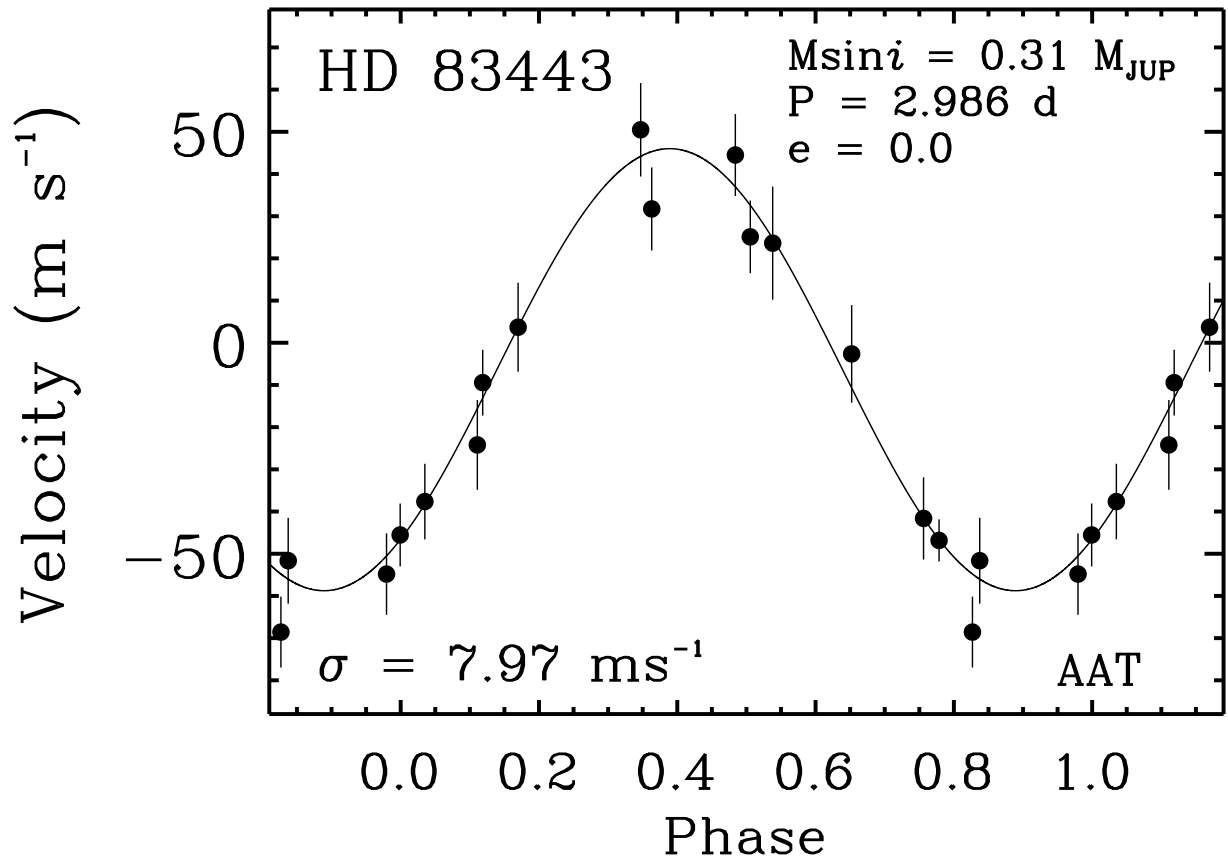}}}}
\caption{Phased Doppler velocities for HD 83443 from the AAT data.
The solid line is the best--fit Keplerian orbit assuming only a single planet.
The period, p = 2.986 d, and semiamplitude, K = 52.4 \ms, are similar
to the CORALIE parameters for the inner planet.  The RMS to the Keplerian
fit, 8 \ms, is consistent with measurement uncertainty.}
\label{AAT_Phased_Velocities}
\end{figure}

\begin{figure}
\centerline{\scalebox{.75}{\rotatebox{90}{\includegraphics{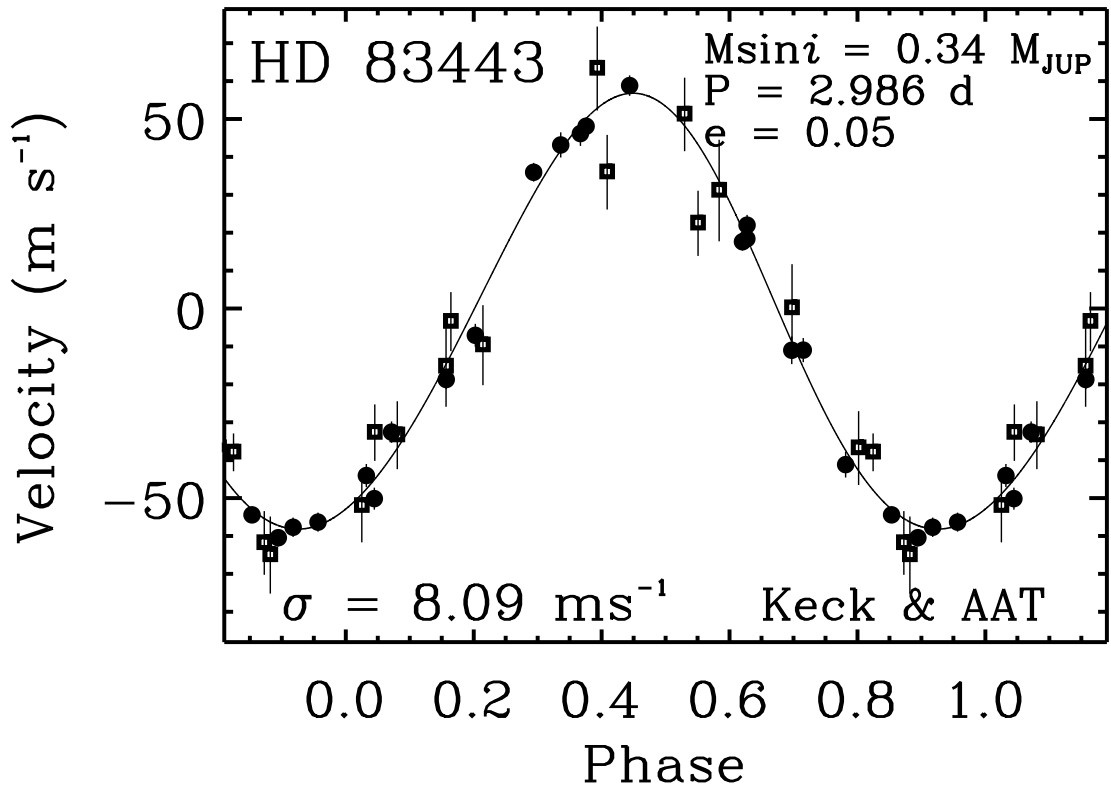}}}}
\caption{Phased Doppler velocities for HD 83443 from the
combined Keck (dots) and AAT (squares) data.  The solid
line is the best--fit Keplerian orbit.  The period,
p = 2.986 d, and semiamplitude, K = 57 \ms, are nearly
identical to the CORALIE parameters for the inner planet.
Within measurement uncertainty, the eccentricity derived
from the Keck--AAT data set is consistent with zero,
similar to other ``51 Peg--like'' planets.}
\label{Keck_AAT_Phased_Velocities}
\end{figure}

\begin{figure}
\centerline{\scalebox{.75}{\rotatebox{90}{\includegraphics{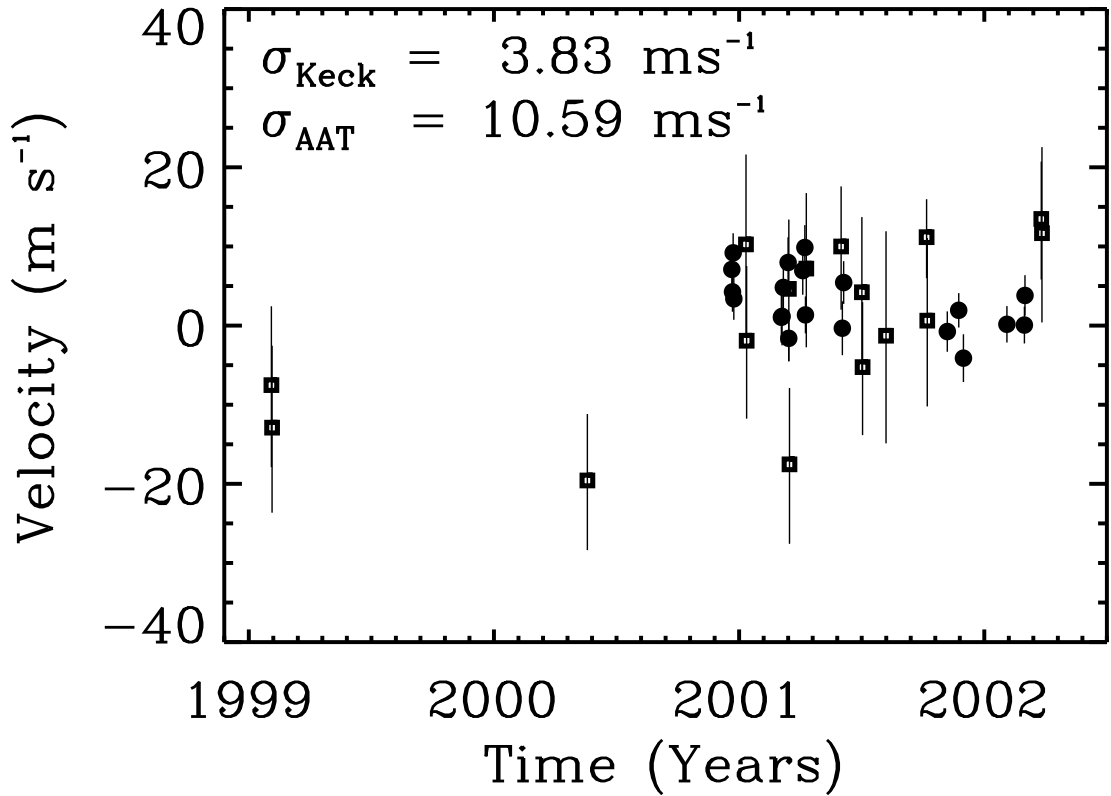}}}}
\caption{Residual velocities from the best--fit single
Keplerian for HD 83443, using the combined Keck (dots) and AAT
(squares) data.  The Keck residuals have an RMS of 3.8 \ms,
consistent with the combined effects of measurement uncertainty
and Doppler jitter.  The AAT residuals have an RMS of 10.6 \ms,
consistent with measurement uncertainty and jitter.}
\label{Keck_AAT_Phased_Velocities}
\end{figure}

\begin{figure}
\centerline{\scalebox{.75}{\rotatebox{90}{\includegraphics{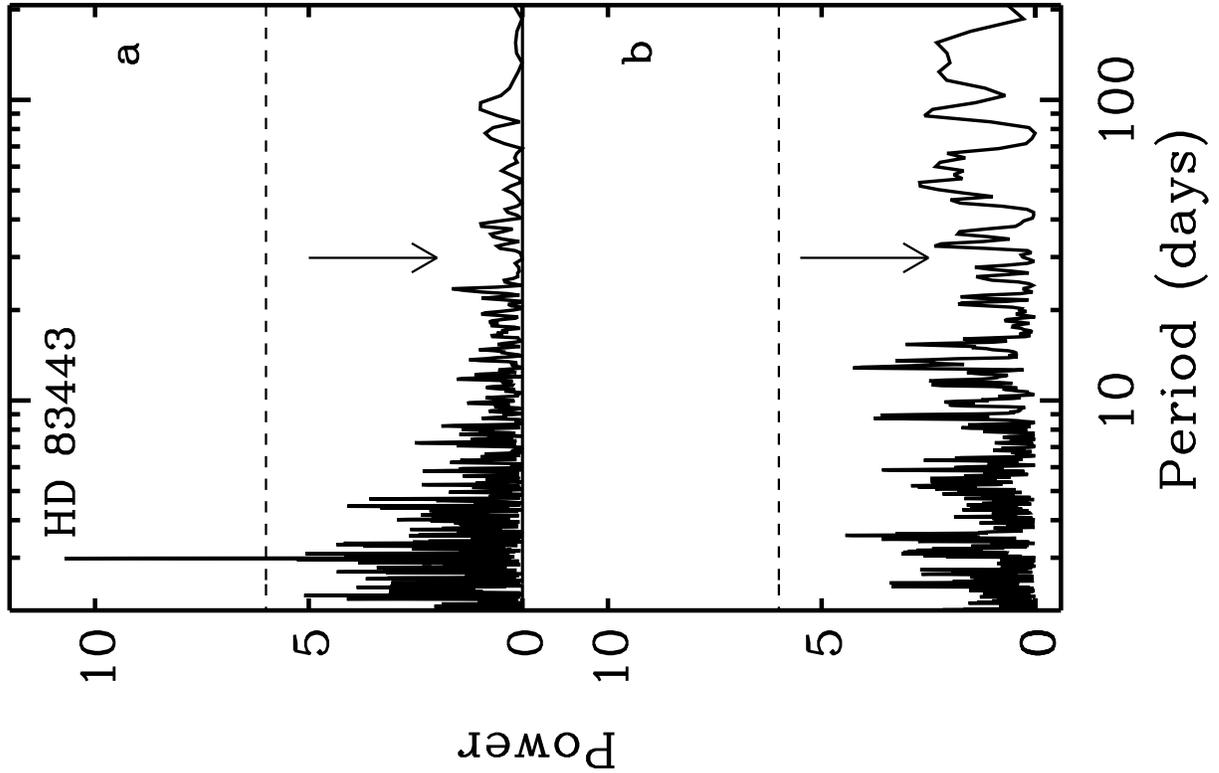}}}}
\caption{Periodogram of HD 83443 Keck velocities. 
a)  Periodogram of measured velocities.  The 2.986 d
periodicity is indicated by the highest periodogram
peak.  The 1\% false alarm level is indicated with the
dotted line.  b)  Periodogram of residual velocities,
after subtracting off the best--fit Keplerian.
No significant periodicities remain after subtracting off the
best--fit single Keplerian.  The arrows indicate 29.83 d,
the purported  period of the outer planet from the
CORALIE data.}
\label{Keck_Periodogram}
\end{figure}

\begin{figure}
\centerline{\scalebox{.75}{\rotatebox{90}{\includegraphics{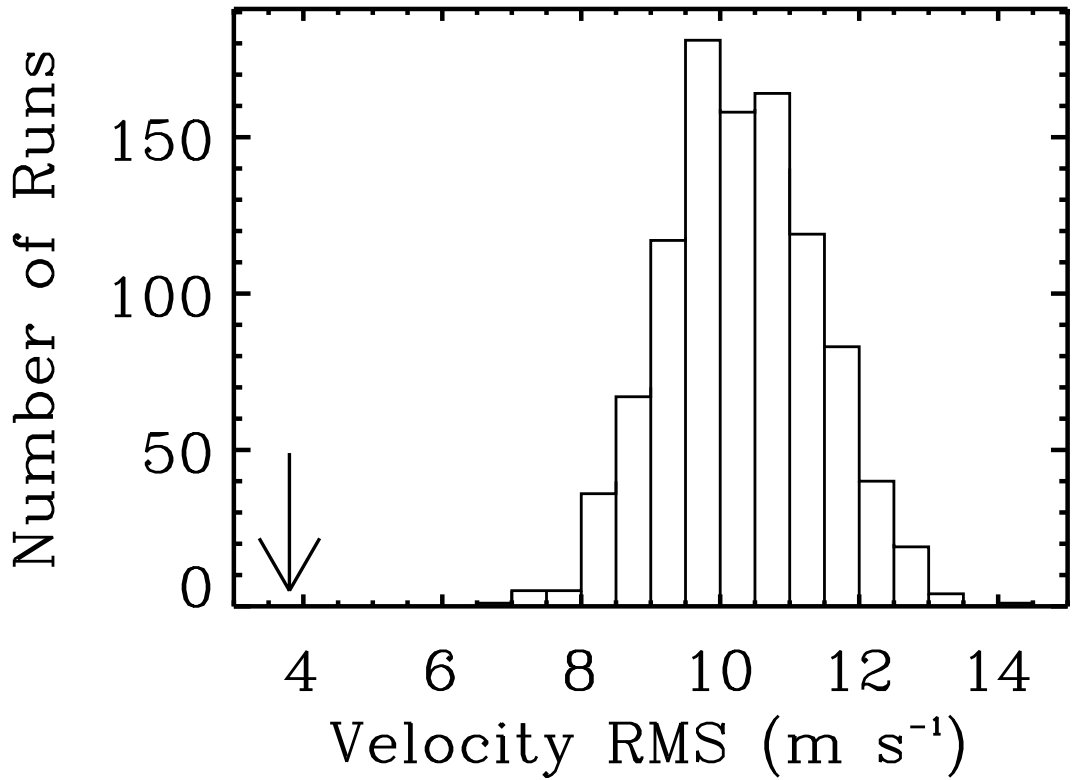}}}}
\caption{Histogram of the RMS of the residuals of a single Keplerian
fit to synthetic velocities that stem from a double--planet system.
One thousand synthetic Doppler velocity sets were constructed, sampled
at the times of the Keck observations, including Gaussian noise.  The
RMS of the residuals to these fits range from 6.7 to 14.0 \ms, with a
median of 10.3 \ms, well above our errors of 3 \ms.  Thus a single
Keplerian model should fail to adequately fit the double--planet
system that was reported.  In contrast, the RMS of the single
Keplerian fit to the actual Keck data yields an RMS of only 
3.8 \ms consistent with noise, indicated by the arrow, suggesting
that the second planet does not exist.}
\label{Histogram}
\end{figure}

\begin{figure}
\centerline{\scalebox{.75}{\rotatebox{90}{\includegraphics{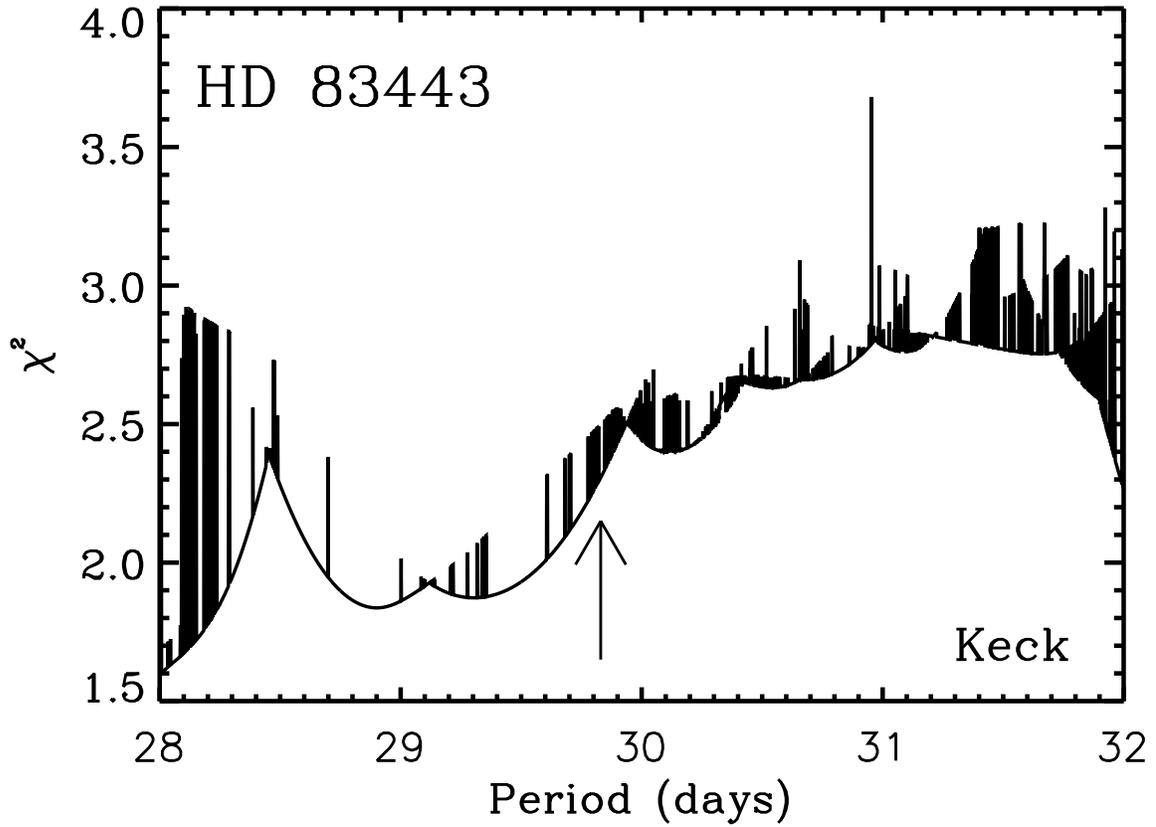}}}}
\caption{Reduced $\chi^2$ as a function of outer planet period for a
2--Keplerian fit to the Keck data.  The period, eccentricity, and
amplitude of the inner planet have been frozen at the CORALIE values, as
have the eccentricity and amplitude of the outer planet.  No minimum
is seen in the reduced $\chi^2$ near 29.83 d, the purported period of
the outer planet.}
\label{Keck_Chisq}
\end{figure}

\begin{figure}
\centerline{\scalebox{.75}{\rotatebox{90}{\includegraphics{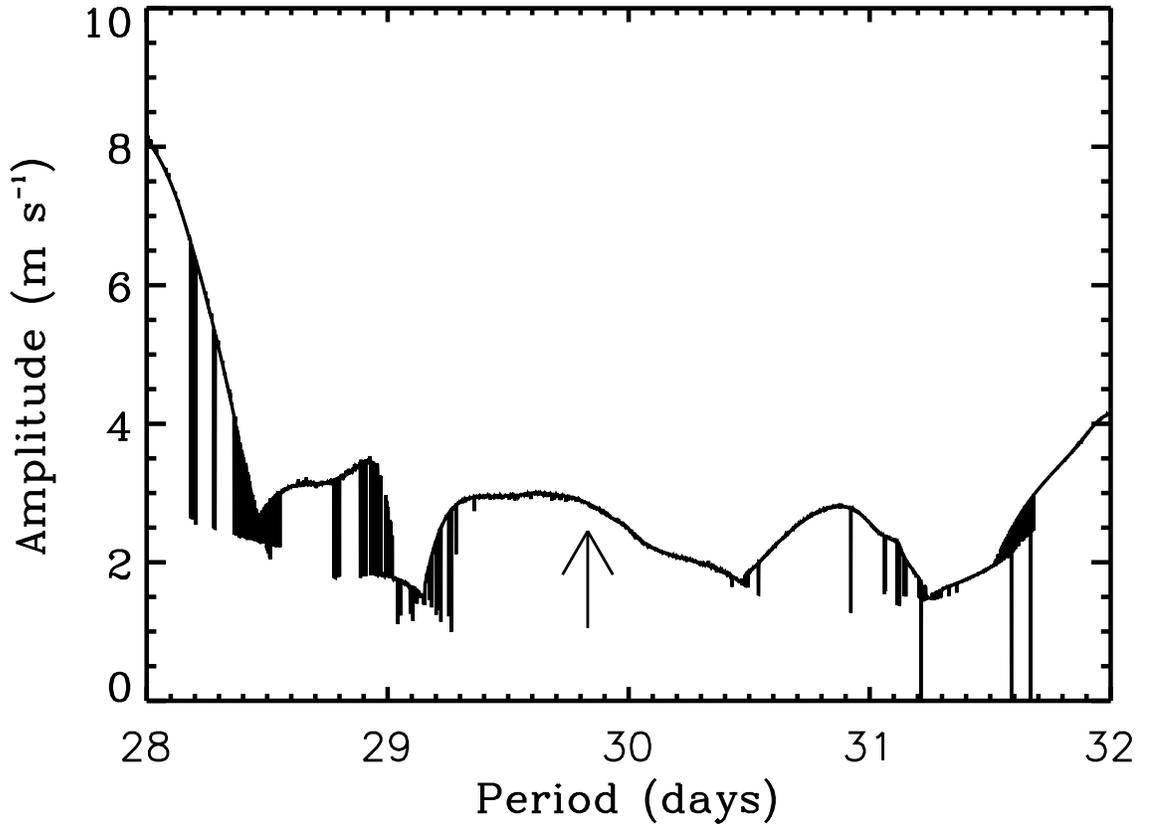}}}}
\caption{Best--fit semiamplitude for the outer planet in a double
Keplerian fit to the Keck data.  The period, eccentricity, and
amplitude of the inner planet have been frozen at the values reported
by CORALIE, as well as the eccentricity of the outer planet.  The Keck
data rule out an outer planet with a semiamplitude greater than 3 \ms
for periods between 29 and 31 days.}
\label{Keck_K}
\end{figure}

\clearpage

\begin{deluxetable}{rrrr}
\tablenum{1}
\tablecaption{Velocities for HD 83443}
\label{vel83443}
\tablewidth{0pt}
\tablehead{
\colhead{JD}           &    \colhead{RV}         & \colhead{error} & \colhead{Tel.} \\
\colhead{($-$2450000)}   &  \colhead{(m s$^{-1}$)} & \colhead{(m s$^{-1}$)} &  \colhead{ } 
}
\startdata
   212.1830  &   -61.0  & 10.2 & AAT \\
   213.1756  &    -5.7  & 10.6 & AAT \\
   682.9088  &    26.5  &  8.6 & AAT \\
   898.0961  &    26.0  &  2.7 & Keck \\
   899.0788  &   -52.3  &  2.5 & Keck \\
   900.0854  &    39.9  &  2.5 & Keck \\
   901.0806  &    22.4  &  2.6 & Keck \\
   919.2047  &     4.1  & 11.6 & AAT \\
   920.1821  &   -48.0  &  9.6 & AAT \\
   971.9566  &    50.2  &  3.2 & Keck \\
   972.9432  &    -7.0  &  3.7 & Keck \\
   974.8502  &    47.2  &  3.3 & Keck \\
   981.9535  &    -7.0  &  3.2 & Keck \\
   982.9366  &   -46.2  &  2.9 & Keck \\
   983.0440  &   -29.4  &  9.0 & AAT \\
   984.0236  &    40.0  &  9.8 & AAT \\
  1003.7982  &   -40.1  &  3.1 & Keck \\
  1006.9015  &   -28.6  &  2.8 & Keck \\
  1007.8096  &    52.1  &  2.3 & Keck \\
  1009.0816  &   -32.8  &  9.7 & AAT \\
  1060.9180  &     0.5  &  7.8 & AAT \\
  1062.7608  &   -37.2  &  3.4 & Keck \\
  1064.7379  &    62.8  &  2.7 & Keck \\
  1091.8643  &    55.2  &  9.7 & AAT \\
  1092.8878  &   -57.8  &  8.4 & AAT \\
  1127.8521  &    35.1  & 13.4 & AAT \\
  1188.2812  &   -33.9  &  5.0 & AAT \\
  1189.2733  &   -11.3  & 10.6 & AAT \\
  1219.1297  &   -14.7  &  2.5 & Keck \\
  1236.1362  &   -50.4  &  2.2 & Keck \\
  1243.1498  &    -3.1  &  3.0 & Keck \\
  1307.9118  &   -56.5  &  2.3 & Keck \\
  1333.9629  &    21.6  &  2.3 & Keck \\
  1334.8507  &   -53.7  &  2.6 & Keck \\
  1359.1162  &   -28.7  &  7.4 & AAT \\
  1360.1546  &    67.3  & 11.1 & AAT \\
\enddata
\end{deluxetable}

\clearpage

\begin{deluxetable}{rlllllllll}
\tablenum{2}
\rotate
\tablecaption{Orbital Parameters}
\label{candid}
\tablewidth{0pt}
\tablehead{
\colhead{Star}  & \colhead{Period} & \colhead{$K$}   & \colhead{$e$} & \colhead{$\omega$}  & \colhead{$T_0$} & \colhead{M$\sin i$} & \colhead{$a$} & \colhead{N$_{\rm obs}$} & \colhead{RMS} \\
\colhead{ }     & \colhead{(days)} & \colhead{(\ms)} &\colhead{ }    & \colhead{(degrees)} & \colhead{(JD-2450000)}  & \colhead{(M$_{J}$)} & {(AU)} & \colhead{ } & \colhead{(m s$^{-1}$)}
}

\startdata
Keck\tablenotemark{a} &      2.98571 (0.001) & 57.0 (4) & 0.059 (0.06)  &  44 (40) & 1876.99 (0.15) & 0.35 & 0.0375  & 20 & 3.81 \\
AAT\tablenotemark{b} &  2.98559 (0.0006)  & 52.9 (5)  & 0.00 & 0 & 1213.8 (0.1) & 0.32 & 0.0375 & 21 & 8.47 \\
Keck--AAT\tablenotemark{c} & 2.98553 (0.0004) & 57.5 (2)   & 0.052 (0.05) &   46 (30) & 1211.24 (0.1) & 0.34 & 0.0375 & 36 & 8.09 \\
CORALIE b & 2.9853 (0.0009)  & 56.1 (1.4) & 0.079 (0.033) &  300 (17) & 1386.50 (0.14) & 0.34 & 0.0380 & 93 & 6 \\
CORALIE c &  29.83 (0.18)    & 13.8  (1)  & 0.42 (0.06)   &  337 (10) & 1569.59 (0.73) & 0.16 & 0.17   & 93 & 6 \\
\enddata
\tablenotetext{a}{Linear slope -5.5 (3) \ms per year.}
\tablenotetext{b}{Forced circular orbit, linear slope +7.4 (3) \ms per year.}
\tablenotetext{c}{Linear slope 0.0 (1) \ms per year.}
\end{deluxetable}

\clearpage

\begin{deluxetable}{rrrllllll}
\tablenum{3}
\tablecaption{Precision Doppler Planets}
\label{candid}
\tablewidth{0pt}
\tablehead{
\colhead{Star} & \colhead{Star} & \colhead{Star} & \colhead{Paper} & \colhead{Date Received} & \colhead{Velocities} & \colhead{Tel.} \\
\colhead{(HD)} & \colhead{(Hipp)} & &  &  & &
} 
\startdata
217014 & 113357 & 51 Peg     & Mayor \& Queloz 1995 & 1995\tablenotemark{a} & N & Elodie \\  
   ... &    ... &    ...     & Marcy et al. 1997    & 1996 Sep 19 & Y & Lick   \\
117176 & 65721  & 70 Vir     & Marcy \& Butler 1996 & 1996 Jan 22 & N & Lick   \\
95128b & 53721b & 47 UMa b   & Butler \& Marcy 1996 & 1996 Feb 15 & N & Lick   \\
   ... &    ... &    ...     & Fischer et al. 2002  & 2001 Mo  Da & Y & Lick   \\
120136 & 67275  & $\tau$ Boo & Butler et al. 1997   & 1996 Aug 12 & N & Lick   \\
75732b & 43587b & 55 Cnc b   & Butler et al. 1997   & 1996 Aug 12 & N & Lick   \\
 9826b &  7513b & $\upsilon$ And b & Butler et al. 1997 & 1996 Aug 12 & N & Lick   \\
   ... &    ... &        ...   & Butler et al. 1999 & 1999 Apr  8 & Y & Lick, AFOE \\
186408 & 96895  & 16 Cyg B   & Cochran et al. 1997  & 1996 Nov 21 & Y & Lick, McDonald \\
143761 &        & HR 5968    & Noyes et al. 1997    & 1997 Apr 18 & N & AFOE \\
       & 113020b & GJ 876 b  & Marcy et al. 1998    & 1998 Jul  7 & N & Lick, Keck \\
       &     ... &    ...    & Delfosse al. 1998    & 1998 Aug 17 & N & Elodie, Coralie \\
       &     ... &    ...    & Marcy et al. 2001b   & 2000 Dec 27 & Y & Lick, Keck \\
187123 & 97336   &           & Butler et al. 1998   & 1998 Sep  6 & N & Keck   \\
   ... &   ...   &           & Vogt et al. 2000     & 1999 Nov 15 & Y & Keck   \\
195019 & 100970  &           & Fischer et al. 1999  & 1998 Oct  8 & Y & Lick   \\
   ... &    ...  &           & Vogt et al. 2000     & 1999 Nov 15 & Y & Keck   \\
217107 & 113421  & HR 8734   & Fischer et al. 1999  & 1998 Oct  8 & Y & Lick   \\
   ... &    ...  &     ...   & Vogt et al. 2000     & 1999 Nov 15 & Y & Keck   \\
   ... &    ...  &     ...   & Naef. et al. 2001a   & 2000 Aug 30 & Y & Coralie \\
210277 & 109378  &           & Marcy et al. 1999    & 1998 Dec 16 & Y & Keck   \\
   ... &    ...  &           & Vogt et al. 2000     & 1999 Nov 15 & Y & Keck   \\
   ... &    ...  &           & Naef. et al. 2001a   & 2000 Aug 30 & Y & Coralie \\
168443b & 89844b &           & Marcy et al. 1999    & 1998 Dec 16 & Y & Keck   \\
    ... &    ... &           & Marcy et al. 2001a   & 2000 Dec 13 & Y & Keck   \\
 9826c &  7513c & $\upsilon$ And c & Butler et al. 1999 & 1999 Apr  8 & Y & Lick, AFOE \\
 9826d &  7513d & $\upsilon$ And d & Butler et al. 1999 & 1999 Apr  8 & Y & Lick, AFOE \\
 13445 & 10138  & GL 86      & Queloz et al. 2000   & 1999 Apr 22 & N & Coralie \\
   ... &   ...  &   ...      & Butler et al. 2001   & 2000 Dec 25 & Y & AAT  \\
 17051 & 12653  & $\iota$ Hor & Kurster et al. 2001 & 1999 Oct 19 & Y & ESO \\
   ... &   ...  &         ... & Naef. et al. 2001a  & 2000 Aug 30 & Y & Coralie \\
   ... &   ...  &         ... & Butler et al. 2001  & 2000 Dec 25 & Y & AAT  \\
 10697 &  8159  &            & Vogt et al. 2000     & 1999 Nov 15 & Y & Keck   \\
37124b & 26381b &            & Vogt et al. 2000     & 1999 Nov 15 & Y & Keck   \\
   ... &    ... &            & Butler et al. 2002   & 2002 May 21 & Y & Keck \\
222582 & 116906 &            & Vogt et al. 2000     & 1999 Nov 15 & Y & Keck   \\
177830 & 93746  &            & Vogt et al. 2000     & 1999 Nov 15 & Y & Keck   \\
134987 & 74500  &            & Vogt et al. 2000     & 1999 Nov 15 & Y & Keck   \\
   ... &   ...  &            & Butler et al. 2001   & 2000 Dec 25 & Y & AAT \\
209458 & 108859 &            & Henry et al. 2000    & 1999 Nov 18 & N & Keck   \\
   ... &    ... &            & Mazeh et al. 2000    & 1999 Dec  3 & N & Elodie, Coralie \\
130332 & 72339  &            & Udry et al. 2000     & 1999 Dec  2 & N & Coralie \\
 75289 & 43177  &            & Udry et al. 2000     & 1999 Dec  2 & N & Coralie \\
   ... &   ...  &            & Butler et al. 2001   & 2000 Dec 25 & Y & AAT  \\
 89744 &  50786 & HR 4067   & Korzennik et al. 2000 & 2000 Jan 20 & N & AFOE, Lick \\
 16141 & 12048  &            & Marcy et al. 2000    & 2000 Mar  6 & Y & Keck   \\
 46375 & 31246  &            & Marcy et al. 2000    & 2000 Mar  6 & Y & Keck   \\
       &        & BD -103166 & Butler et al. 2000   & 2000 Apr 21 & Y & Keck   \\
 52265 & 33719  & HR 2622    & Butler et al. 2000   & 2000 Apr 21 & Y & Keck   \\
   ... &   ...  & ...        & Naef. et al. 2001a   & 2000 Aug 30 & Y & Coralie \\
12661b & 9683b  &            & Fischer et al. 2001  & 2000 Jul 19 & Y & Lick, Keck \\
 92788 & 52409  &            & Fischer et al. 2001  & 2000 Jul 19 & Y & Lick, Keck \\
38529b & 27253b &            & Fischer et al. 2001  & 2000 Jul 19 & Y & Lick, Keck \\
 22049 & 16537  & $\epsilon$ Eri & Hatzes et al. 2000   & 2000 Aug 22 & Y & McDonald, CFHT, ESO \\
169830 & 90485  &            & Naef. et al. 2001a   & 2000 Aug 30 & Y & Coralie \\
  1237 &  1292  & GJ 3021    & Naef. et al. 2001a   & 2000 Aug 30 & Y & Coralie \\
179949 & 94645  &            & Tinney et al. 2001   & 2000 Oct 11 & Y & AAT, Keck \\
160691 & 86796  & HR 6585    & Butler et al. 2001   & 2000 Dec 25 & Y & AAT  \\
 27442 & 19921  & HR 1355    & Butler et al. 2001   & 2000 Dec 25 & Y & AAT  \\
       & 113020c & GJ 876 c  & Marcy et al. 2001b   & 2000 Dec 27 & Y & Lick, Keck \\
 80606 & 45982  &            & Naef et al. 2001b    & 2001 May 29 & Y & Coralie \\
95128c & 53721c & 47 UMa c   & Fischer et al. 2002  & 2001 Jun 29 & Y & Lick \\
28185  & 20723  &            & Santos et al. 2001   & 2001 Jul 30 & Y & Coralie \\
213240 & 111143 &            & Santos et al. 2001   & 2001 Jul 30 & Y & Coralie \\
 4208  & 3479   &            & Vogt et al. 2002     & 2001 Oct 16 & Y & Keck \\
114783 & 64467  &            & Vogt et al. 2002     & 2001 Oct 16 & Y & Keck \\
 4203  & 3502   &            & Vogt et al. 2002     & 2001 Oct 16 & Y & Keck \\
 68988 & 40687  &            & Vogt et al. 2002     & 2001 Oct 16 & Y & Keck \\
 33636 & 24205  &            & Vogt et al. 2002     & 2001 Oct 16 & Y & Keck \\
   142 &   522  & HR 6       & Tinney et al. 2002   & 2001 Nov 12 & Y & AAT  \\
 23079 & 17096  &            & Tinney et al. 2002   & 2001 Nov 12 & Y & AAT  \\
 39091 & 26394  & HR 2022    & Jones et al. 2002    & 2001 Nov 29 & Y & AAT  \\
108147 & 60644  &            & Pepe et al. 2002     & 2002 Feb 26 & Y & Coralie \\
168746 & 90004  &            & Pepe et al. 2002     & 2002 Feb 26 & Y & Coralie \\
141937 & 77740  &            & Udry et al. 2002     & 2002 Feb 26 & Y & Coralie \\
137759 & 75458  & $\iota$ Dra & Frink et al. 2002    & 2002 Mar 21& Y & Lick \\
83443b & 47202  &            & This Paper           & 2002 Apr 8 & Y & AAT, Keck\\
\enddata
\tablenotetext{a}{Note: Nature does not publish ``Date Received''.}
\end{deluxetable}

\clearpage


\clearpage

\begin{thebibliography}{}
\parsep 0pt
\itemsep -3pt

\bibitem[Butler { et~al.} 1996]{BuMaWi96} Butler, R.~P., Marcy, G.~W.,
Williams, E., McCarthy, C., Dosanjh, P., \& Vogt, S.~S. 1996,
\newblock { PASP, } {108}, 500

\bibitem[Butler \& Marcy 1996]{BuMa96}
Butler, R.~P. \&  Marcy, G.~W. 1996,
\newblock {  ApJ, } {464}, L153.

\bibitem[Butler {  et~al.} 1997]{BuMaWi97}
Butler, R.~P., Marcy, G.~W., Williams, E., Hauser, H., \& Shirts, P.  1997,
\newblock {  ApJ, } {474}, L115

\bibitem[Butler \& Marcy 1997]{BuMa97}
Butler, R.~P. \& Marcy, G.~W. 1997,
``The Near Term Future of Extrasolar Planet Searches'',
Brown Dwarfs and Extrasolar Planets, held on Tenerife, 17-21 March 1997,
ed. R. Rebolo, E.L. Martin, and M.R. Zapatero Osorio,
ASP Conference Series, Vol. 134, p. 162. 

\bibitem[Butler { et~al.} 1998]{BuMa98}
Butler, R.~P., Marcy, G.~W., Vogt, S.~S. \& Apps, K. 1998,
\newblock { PASP, } {110}, 1389.

\bibitem[Butler {  et~al.} 1999]{BuMaFi99}
Butler, R.~P., Marcy, G.~W., Fischer, D.~A., Brown, T.~M.,
Contos, A.~R., Korzennik, S.~G., Nisenson, P. \& Noyes, R.~W. 1999,
\newblock {  ApJ, } {526}, 916.

\bibitem[Butler { et~al.} 2000]{BuMa00}
Butler, R.~P., Vogt, S.~S., Marcy, G.~W., Fischer, D.~A.,
Henry, G.~W. \& Apps, K. 2000,
\newblock { ApJ, } {545}, 504.

\bibitem[Butler { et~al.} 2001]{BuMa01}
Butler, R.~P., Tinney, C.~G., Marcy, G.~W., Jones, H.~R.~A.,
Penny, A.~J.  \& Apps, K. 2001,
\newblock { ApJ, } {555}, 410.

\bibitem[Chiang { et~al.} 2002]{Chiang02}
Chiang, E. I., Fischer, D., Thommes, E. 2002, ApJ, 564L, 105

\bibitem[Cochran { et~al.} 1997]{Coch97}
Cochran, W.~D., Hatzes, A.~P., Butler, R.~P \& Marcy, G.~W. 1997,
\newblock { ApJ, } {483}, 457.


\bibitem[Delfosse et al. 1998]{Del98}
Delfosse, X., Forveille, T., Mayor, M., Perrier, C.,
Naef, D. \& Queloz, D. 1998
\newblock {  A\&A, } {338}, L67.

\bibitem[Diego {  et~al.} 1991]{diego:90}
Diego, F., Charalambous, A., Fish, A.~C., \& Walker, D.~D. 1990,
Proc. Soc. Photo-Opt. Instr. Eng., 1235, 562

\bibitem[ESA 1997]{ESA97}
ESA 1997, The Hipparcos and Tycho Catalogues (ESA SP-1200).

\bibitem[Fischer {  et~al.} 1999]{FiMa99}
Fischer, D.~A., Marcy, G.~W., Butler, R.P., Vogt, S.~S.\& Apps, K. 1999,
\newblock {  PASP, } {111}, 50 

\bibitem[Fischer {  et~al.} 2001]{Fischer01}
Fischer, D.~A., Marcy, G.~W., Butler, R.P., Vogt, S.~S.,
Frank. S. \& Apps, K. 2001,
\newblock {  ApJ, } {551}, 1107

\bibitem[Fischer {  et~al.} 2002]{Fisch02}
Fischer, D.~A., Marcy, G.~W., Butler, R.P., Laughlin, G. \& Vogt, S.~S. 2002,
\newblock {  ApJ, } 564, 1028.

\bibitem[Gilliland et al. 1987]{Gill87}
Gilliland, R.~L. \& Baliunas, S.~L. 1987,
\newblock {  ApJ, } 314, 766

\bibitem[Hatzes{  et~al.} 2000]{Hatzes00}
Hatzes, A.~P., Cochran, W.~D, McArthur, B., Baliunas, S.~L.,
Walker, G.~A.~H., Campbell, B., Irwin, A.~W., Yang, S.,
Kurster, M., Endl, M., Els, S., Butler, R.~P. \& Marcy, G.~W. 2000,
\newblock { ApJ, } {544}, L145.

\bibitem[Henry {  et~al.} 2000]{Henry00}
Henry, G.~W., Marcy, G.~W., Butler, R.~P. \& Vogt, S.~S. 2000,
\newblock { ApJ, } {529}, L45.

\bibitem[Jones et al. 2002]{jones:02}
Jones, H.~R.~A., Butler, R.~P., Tinney, C.~G., Marcy, G.~W.,
Penny, A.~J., McCarthy, C., Carter, B.~D. \& Apps, K. 2002,
\newblock {  MNRAS, }, in press.

\bibitem[Korzennik et al. 2000]{korzen00}
Korzennik, S.~G., Brown, T.~M., Fischer, D.~A.,
Nisenson, P. \& Noyes, R.~W.  2000,
\newblock {  ApJ, } {533}, L147.

\bibitem[Kurster 2000]{kurst00}
Kurster, M., Endl, M., Els, S., Hatzes, A.~P., Cochran, W.~D.,
Dobereiner, S. \& Dennerl, K. 2000,
\newblock {  A\&A, } 353, L33.

\bibitem[Laughlin & Chambers 2001]{Laug01}
Laughlin, G. \& Chambers, J~E. 2001,
\newblock {  ApJ, } {551}, L109.

\bibitem[Lee \& Peale 2002a]{Lee02a}
Lee, M.H., Peale, S.J. 2002a, ApJ, 567, 596

\bibitem[Lee \& Peale 2002b]{Lee02b}
Lee, M.H., Peale, S.J. 2002b, personal communication

\bibitem[Lissauer & Rivera 2001]{Liss01}
Lissauer, J.~J. \& Rivera, E.~J. 2001,
\newblock {  ApJ, } {554}, 1141.


\bibitem[Marcy \& Butler 1992]{MaBu92}
Marcy, G.~W. \&  Butler, R.~P. 1992,
\newblock {  PASP, } {104}, 270.

\bibitem[Marcy \& Butler 1996]{MaBu96}
Marcy, G.~W. \& Butler, R.~P. 1996,
\newblock {  ApJ, } {464}, L151.

\bibitem[Marcy { et~al.} 1997]{Marcy97}
Marcy, G.~W., Butler, R.~P., Williams, E., Bildsten, L.,
Graham, J.~R., Ghez A.~M., Jernigan J.~G. 1997,
\newblock {  ApJ, } {481}, 926.

\bibitem[Marcy \& Butler 2000]{MaBu00}
Marcy, G.~W. \&  Butler, R.~P. 2000,
\newblock {  PASP, } {112}, 137.

\bibitem[Marcy { et~al.} 2000]{MaBuVo00}
Marcy, G.~W., Butler, R.~P., Vogt, S.~S. 2000,
\newblock {  ApJ, } {536}, L43.

\bibitem[Marcy { et~al.} 1997]{MaBuWi97} 
Marcy, G.~W., Butler, R.~P., Williams, E., Bildsten, L.,
Graham, J.~R., Ghez, A., \& Jernigan, G. 1997,
\newblock { ApJ, } 481, 926.

\bibitem[Marcy { et~al.} 1998]{Marcy98} 
Marcy, G.~W., Butler, R.~P., Vogt, S.~S.,
Fischer, D.~A. \& Lissauer, J.~J 1998,
\newblock { ApJ, } 505, L147.

\bibitem[Marcy { et~al.} 1999]{Marcy99} 
Marcy, G.~W., Butler, R.~P., Vogt, S.~S.,
Fischer, D.~A. \& Liu, M.~C. 1999,
\newblock { ApJ, } 520, 239.

\bibitem[Marcy {  et~al.} 2001a]{mbv2001a}
Marcy, G.~W., Butler, R.~P., Vogt, S.~S., Liu, M.~C.,
Laughlin, G.~P., Apps, K., Graham, J.~R., Lloyd, J.,
Luhman, K.~L. \& Jaywardhana, R. 2001a,
\newblock { ApJ, } 555, 418.

\bibitem[Marcy {  et~al.} 2001b]{mbv2001b}
Marcy, G.~W., Butler, R.~P., Fischer, D.~A.,
Vogt, S.~S., Lissauer, J.~J. \& Rivera, E.~J. 2001b,
\newblock { ApJ, } 556, 296.

\bibitem[Mayor \& Queloz 1995]{Mayor95}
Mayor, M. \& Queloz, D. 1995, Nature,378,355

\bibitem[Mayor et al. 2000]{Mayor00}
Mayor, M., Naef, D., Pepe, F., Queloz, D., Santos, N.~C.,
Udry, S. \& Burnet, M. 2000,
``HD 83443: A System with Two Saturns'',
Planetary Systems in the Universe, held at the
IAU General Assembly, Manchester UK, 7 August 2000,
ed. A.J. Penny, P. Artymowicz, A.M. Lagrange, and S.S. Russell,
ASP Conference Series, in press. 

\bibitem[Mazeh et al. 2000]{Mazeh00}
Mazeh, T., Naef, D., Torres, G., Latham, D.~W., Mayor, M.,
Beuzit, J.~L., Brown, T.~M., Buchhave, L., Burnet, M.,
Carney, B.~W., Charbonneau, D., Drukier, G.~A., Laird, J.~B.,
Pepe, F., Perrier, C., Queloz, D., Santos, N.~C., Sivan, J.~P,
Udry, S., Zucker, S. 2000,
\newblock { ApJ, } 532, L55.

\bibitem[Naef et al. 2001]{Naef2001}
Naef, D., Mayor, M., Pepe, F., Queloz, D., Santos, N.~C.,
Udry, S. \& Burnet, M. 2001a,
\newblock {  A\&A, } 375, 205.

\bibitem[Naef et al. 2001]{Naef2001}
Naef, D., Latham, D.~W., Mayor, M., Mazeh, T., Beuzit, J.~L.,
Drukier, G.~A., Perrier-Bellet, C., Queloz, D., Sivan, J.~P.,
Torres, G., Udry, S. \& Zucker, S.  2001b,
\newblock {  A\&A, } 375, L27.

\bibitem[Noyes et al. 1997]{noyes:97}
Noyes, R.~W., Jha, S., Korzennik, S.~G., Krockenberger, M.,
Nisenson, P., Brown, T.~M., Kennelly, E.~J. \& Horner, S.~D. 1997,
\newblock {  ApJ, } 483, L111.

\bibitem[Pepe et al. 2002]{Pepe2002}
Pepe, F., Mayor, M., Galland, D., Queloz, D., 
Santos, N.~C., Udry, S. \& Burnet, M.  2002,
\newblock {  A\&A, } submitted.

\bibitem[Perryman et al. 1997]{Perry97}
Perryman, M.~A.~C., et al. 1997, { A\&A, } 323, L49. The Hipparcos Catalog

\bibitem[Queloz et al. 2000]{Quel2000}
Queloz, D., Mayor, M., Weber, L., Blecha, A., Burnet, M.,
Confino, B., Naef, D., Pepe, F., Santos, N.~C. \& Udry, S. 2000a,
\newblock {  A\&A, } 354, 99.

\bibitem[Queloz et al. 2000]{Queloz00}
Queloz, D., Mayor, M., Naef, D., Pepe, F., Santos, N.~C.,
Udry, S. \& Burnet, M. 2000b,
``4 Jovian Extrasolar Planets Detected with CORALIE'',
Planetary Systems in the Universe, held at the
IAU General Assembly, Manchester UK, 7 August 2000b,
ed. A.J. Penny, P. Artymowicz, A.M. Lagrange, and S.S. Russell,
ASP Conference Series, in press. 

\bibitem[Rivera & Lissauer 2001]{Rivera01}
Rivera, E.~J. \&  Lissauer, J.~J. 2001, 
\newblock {  ApJ, } 558, 392.

\bibitem[Saar {  et~al.} 1998]{SaBuMa98}
Saar, S.~H., Butler, R.~P., \& Marcy, G.~W. 1998,
\newblock {  ApJ, } 498, L153.

\bibitem[Saar \& Fischer 2000]{SaFi00}
Saar, S.~H. \& Fischer, D.~A. 2000,
\newblock {  ApJ, } 534, L105.

\bibitem[Santos {  et~al.} 2000]{Santos00}
Santos, N.~C., Israelian, G. \& Mayor, M. 2000a
\newblock {  A\&A }, 363, 228.

\bibitem[Santos {  et~al.} 2000]{Santos00}
Santos, N.~C., Mayor, M., Naef, D., Pepe, F., Queloz, D.,
Udry S., \& Blecha, A 2000b,
\newblock {  A\&A }, 356, 599.

\bibitem[Santos {  et~al.} 2001]{Santos01}
Santos, N.~C., Israelian, G., Mayor, M.  2001, A\&A, 373, 1019

\bibitem[Scargle 1982]{Scarg82}
Scargle, J.~D. 1982,
\newblock {  ApJ, } 263, 835.

\bibitem[Sivan et al. 2000]{Sivan00}
Sivan, J.~P., Mayor, M., Naef, D., Queloz, D.,
Udry, S., Perrier--Bellet, C., \& Benzit, J.~L. 2000,
``A Planetary Companion to HD 190228'',
Planetary Systems in the Universe, held at the
IAU General Assembly, Manchester UK, 7 August 2000b,
ed. A.J. Penny, P. Artymowicz, A.M. Lagrange, and S.S. Russell,
ASP Conference Series, in press. 

\bibitem[Snellgrove{  et~al.} 2001]{Snell01}
Snellgrove M., Papaloizou,J.C.B., Nelson R. 2001, A\&A, 374, 1092

\bibitem[Tinney et al. 2001]{tinney:01}
Tinney, C.~G., Butler, R.~P., Marcy, G.~W., Jones, H.~R.~A.,
Penny, A.~J., Vogt, S.~S., Henry, G.~W. 2001,
\newblock {  ApJ, }, 551, 507.

\bibitem[Tinney et al. 2002]{tinney:02}
Tinney, C.~G., Butler, R.~P., Marcy, G.~W., Jones, H.~R.~A.,
Penny, A.~J., McCarthy, C. \& Carter, B.~D. 2002,
\newblock {  ApJ, }, in press.

\bibitem[Udry et al. 2000]{udry:00}
Udry, S., Mayor, M., Naef, D., Pepe, F., Queloz, D., Santos, N.,
Burnet, M., Confino, B. \& Melo, C. 2000, { A\&A, } 356, 590.

\bibitem[Udry et al. 2002]{udry:02}
Udry, S., Mayor, M., Naef, D., Pepe, F., Queloz, D.,
Santos, N. \& Burnet, M. 2002, { A\&A, } submitted.

\bibitem[Vogt 1987]{vogt:87}
Vogt, S.~S. 1987, { PASP, } 99, 1214.

\bibitem[Vogt et al. 1994]{vogt:94}
Vogt, S.~S. et al. 1994, Proc. Soc. Photo-Opt. Instr. Eng., 2198, 362

\bibitem[Vogt {  et~al.} 2000]{vogt:00}
Vogt, S.~S., Marcy, G.~W., Butler, R.~P. \& Apps, K. 2000,
\newblock { ApJ, } {536}, 902.

\bibitem[Vogt {  et~al.} 2002]{vogt:02}
Vogt, S.~S., Butler, R.~P., Marcy, G.~W., Fischer, D.~A.,
Pourbaix, D., Apps, K. \& Laughlin, G. 2002,
\newblock { ApJ, } in press.

\bibitem[Wu \& Goldreicn 2002]{wu:02}
Wu,Y. \& Goldreich,P. 2002, ApJ, 564, 1024.

\end{thebibliography}
\end{document}